\documentclass[12pt]{article}
\usepackage{graphicx,color}

\makeatletter
\renewcommand{\p@subsection}{}
\newbox\tempboxa
\newdimen\captionboxsubcount 
\def\capsize#1{\captionboxsubcount=#1pt}
\newdimen\captionboxsub
\captionboxsub=\hsize \advance\captionboxsub by -\captionboxsubcount
\advance\captionboxsub by -\captionboxsubcount
\long\def\@makecaption#1#2{
 \setbox\@tempboxa\hbox{\footnotesize #1: #2}
 \ifdim \wd\@tempboxa >\captionboxsub 
\rightskip=\captionboxsubcount \leftskip=\captionboxsubcount 
  \footnotesize #1: #2 
\else \hbox to\hsize{\hfil\box\@tempboxa\hfil} 
 \fi}
\makeatother
\capsize{30}

\newcommand{\Slash}[1]{\ooalign%
{\hfil\rotatebox{30}{\underline{\hspace*{0.6cm}}}\hfil\crcr$#1$}}

\begin{document}

\renewcommand{\thefootnote}{\fnsymbol{footnote}}

\begin{center}
\Large\bf
Thermal Dilepton Production Rate
from Dropping $\rho$
in the Vector Manifestation~\footnote{%
Talk given by M.~Harada at 
Mini-Workshop ``Strongly Coupled Quark-Gluon Plasma: 
SPS, RHIC and LHC" (16-18 February 2007, Nagoya, Japan).
This is based on the work done in Ref.~\cite{HS:HLSDL}.
}
\end{center}

\begin{center}
{\large Masayasu Harada$^{\rm(a)}$ and Chihiro Sasaki$^{\rm(b)}$}
\end{center}

\begin{center}
$^{\rm(a)}${\it
Department of Physics, Nagoya University,
Nagoya, 464-8602, Japan} \\
$^{\rm(b)}${\it
Physik-Department, Technische
Universit\"at M\"unchen, D-85747 Garching, Germany}
\end{center}

\begin{abstract}
In this write-up we summarize main points of 
our recent analysis on the thermal dilepton
production rate from the
dropping $\rho$ based on the vector manifestation (VM).
In the analysis, we studied the
effect of the strong violation of the vector dominance (VD),
which is predicted by the VM,
and showed that
the effect of the violation of the VD
substantially suppresses the dilepton production rate
compared with the one predicted by assuming the VD together 
with the dropping $\rho$.
\end{abstract}

\setcounter{footnote}{0}

\renewcommand{\thefootnote}{\#\arabic{footnote}}

%%%%%%%%%%%%%%%%%%%%%%%%%%%%%%%%%%%%%%%%%%%%%%%%%%%%%%%%%%%%%
%%%%%%%%%%%%%%%%%%%%%%%%%%%%%%%%%%%%%%%%%%%%%%%%%%%%%%%%%%%%%%

\section{Introduction}
\label{sec:int}

Changes of hadron properties are indications of chiral symmetry 
restoration occurring in hot and/or dense QCD and have been 
explored using various effective chiral approaches~\cite{rest,RW}.
An enhancement of dielectron mass spectra below
the $\rho / \omega$ resonance was first observed at CERN 
SPS~\cite{ceres} and it is an indication of the medium 
modification of the vector mesons.
The vector meson mass in matter still remains an open issue.
Although there are several scenarios
like collisional broadening due to interactions with the
surrounding hot/dense medium~\cite{RW}, 
and dropping $\rho$ meson mass associated with chiral symmetry 
restoration~\cite{BR-scaling,HY:VM},
no conclusive distinction between them has been done.
Indeed, the in-medium modification carried by $\rho$ near the
critical point may not be apparent in the final yield depending 
on the fireball evolution under the CERN SPS condition~\cite{virial}.
However, 
dropping masses of hadrons following the Brown-Rho (BR)
scaling~\cite{BR-scaling} can be one of the most prominent 
candidates of the strong signal of melting quark condensate 
$\langle\bar{q}q\rangle$ which is an order parameter of 
spontaneous chiral symmetry breaking,
if the signal is not washed out through the evolution,
especially at RHIC.

The vector manifestation (VM)~\cite{HY:VM} 
is a novel pattern 
of the Wigner realization of chiral symmetry in which
the $\rho$ meson becomes massless degenerate with the
pion at the chiral phase transition point.
The VM is formulated~\cite{HY:PRep,HS:VM,HKR:VM,Sasaki:D} 
in the effective field theory based
on the hidden local symmetry (HLS)~\cite{BKUYY,BKY:PRep}.
The VM gives
a field theoretical description of the dropping $\rho$ mass,
which is protected by the existence of the
fixed point (VM fixed point).

The dropping mass is supported by
the mass shift of the $\omega$ meson in nuclei measured by
the KEK-PS E325 Experiment~\cite{KEK-PS} and
the CBELSA/TAPS Collaboration~\cite{trnka}
and also that of
the $\rho$ meson observed in the STAR experiment~\cite{SB:STAR}.
Recently NA60 Collaboration has provided data for the dimuon 
spectrum~\cite{NA60} and it seems difficult to explain the data
by a naive dropping $\rho$~\cite{NA60:2}.
However, there are still several ambiguities which are not
considered~\cite{Brown:2005ka-kb,HR,SG}.
Especially, the strong violation of the vector dominance (VD),
which is one of the significant predictions of the VM~\cite{HS:VD},
plays an important role~\cite{Brown:2005ka-kb} 
to explain the data.

In Ref.~\cite{HS:HLSDL},
we studied the dilepton production rate
from the dropping $\rho$ based on the VM
using the HLS theory at finite temperature.
We paid a special attention to 
the effect of the violation of the vector dominance
(indicated by ``$\Slash{\rm VD}$'') 
which is due to the
{\it intrinsic temperature effects} of the parameters
introduced through the matching to QCD in the Wilsonian
sense combined with the renormalization group equations (RGEs).
We made a comparison of the dilepton production rates
predicted by the VM with the ones by the dropping $\rho$
with the assumption of the vector dominance (VD).
The result shows that the effect of the $\Slash{\rm VD}$
substantially suppresses the dilepton production rate
compared with the one predicted by assuming the VD together 
with the dropping $\rho$.

This write-up is organized as follows:
In section~\ref{sec:VM} we explain what the VM is.
Section~\ref{sec:DS} is a main part in which
we show the form factor and dilepton production rate.
A brief summary and discussions are given in section~\ref{sec:sum}.

%%%%%%%%%%%%%%%%%%%%%%%%%%%%%%%%%%%%%%%%%%%%%%%%%%%%%%%%%%%%
%%%%%%%%%%%%%%%%%%%%%%%%%%%%%%%%%%%%%%%%%%%%%%%%%%%%%%%%%%%%%%

\setcounter{equation}{0}

\section{Hidden Local Symmetry and Vector Manifestation}
\label{sec:VM}

The vector manifestation (VM) was first proposed in
Ref.~\cite{HY:VM} as a novel manifestation of Wigner 
realization of
chiral symmetry where the vector meson $\rho$ becomes massless at the
chiral phase transition point. 
Accordingly, the (longitudinal) $\rho$ becomes the chiral partner of
the Nambu-Goldstone (NG) boson $\pi$.
The VM is characterized by
\begin{equation}
\mbox{(VM)} \qquad
f_\pi^2 \rightarrow 0 \ , \quad
m_\rho^2 \rightarrow m_\pi^2 = 0 \ , \quad
f_\rho^2 / f_\pi^2 \rightarrow 1 \ ,
\label{VM def}
\end{equation}
where $f_\rho$ is the decay constant of 
(longitudinal) $\rho$ at $\rho$ on-shell.
This is completely different from 
the conventional picture based
on the linear sigma model 
where the scalar meson $S$ becomes massless
degenerate with $\pi$ as the chiral partner:
\begin{equation}
\mbox{(GL)} \qquad
f_\pi^2 \rightarrow 0 \ , \quad
m_S^2 \rightarrow m_\pi^2 = 0 \ .
\label{GL def}
\end{equation}
In Ref.~\cite{HY:PRep}
this was called GL manifestation after the
effective theory of Ginzburg--Landau or Gell-Mann--Levy.

We first consider 
the representations of 
the following zero helicity ($\lambda=0$) states
under
$\mbox{SU(3)}_{\rm L}\times\mbox{SU(3)}_{\rm R}$;
the $\pi$, the (longitudinal) $\rho$, the (longitudinal) axial-vector
meson denoted by $A_1$ ($a_1$ meson and its flavor partners)
and the scalar meson denoted by $S$.
The $\pi$ and the longitudinal $A_1$ 
are admixture of $(8\,,\,1) \oplus(1\,,\,8)$ and 
$(3\,,\,3^*)\oplus(3^*\,,\,3)$
since the symmetry is spontaneously
broken~\cite{Gilman-Harari-Weinberg}:
\begin{eqnarray}
\vert \pi\rangle &=&
\vert (3\,,\,3^*)\oplus (3^*\,,\,3) \rangle \sin\psi
+
\vert(8\,,\,1)\oplus (1\,,\,8)\rangle  \cos\psi
\ ,
\nonumber
\\
\vert A_1(\lambda=0)\rangle &=&
\vert (3\,,\,3^*)\oplus (3^*\,,\,3) \rangle \cos\psi 
- \vert(8\,,\,1)\oplus (1\,,\,8)\rangle  \sin\psi
\ ,
\label{mix pi a}
\end{eqnarray}
where the experimental value of the mixing angle $\psi$ is 
given by approximately 
$\psi=\pi/4$~\cite{Gilman-Harari-Weinberg}.  
On the other hand, the longitudinal $\rho$
belongs to pure $(8\,,\,1)\oplus (1\,,\,8)$
and the scalar meson to 
pure $(3\,,\,3^*)\oplus (3^*\,,\,3)$:
\begin{eqnarray}
\vert \rho(\lambda=0)\rangle &=&
\vert(8\,,\,1)\oplus (1\,,\,8)\rangle  
\ ,
\nonumber
\\
\vert S\rangle &=&
\vert (3\,,\,3^*)\oplus (3^*\,,\,3) \rangle 
\ .
\label{rhos}
\end{eqnarray}

When the chiral symmetry is restored at the
phase transition point, 
it is natural to expect that
the chiral representations coincide with the mass eigenstates:
The representation mixing is dissolved.
{}From Eq.~(\ref{mix pi a}) one can easily see
that
there are two ways to express the representations in the
Wigner phase of chiral symmetry:
The conventional GL manifestation
corresponds to 
the limit $\psi \rightarrow \pi/2$ in which
$\pi$ is in the representation
of pure $(3\,,\,3^*)\oplus(3^*\,,\,3)$ 
together with the scalar meson, 
both being the chiral partners:
\begin{eqnarray}
\mbox{(GL)}
\qquad
\left\{
\begin{array}{rcl}
\vert \pi\rangle\,, \vert S\rangle
 &\rightarrow& 
\vert  (3\,,\,3^\ast)\oplus(3^\ast\,,\,3)\rangle\ ,
\\
\vert \rho (\lambda=0) \rangle \,,
\vert A_1(\lambda=0)\rangle  &\rightarrow&
\vert(8\,,\,1) \oplus (1\,,\,8)\rangle\ .
\end{array}\right.
\end{eqnarray}
On the other hand, the VM corresponds 
to the limit $\psi\rightarrow 0$ in which the $A_1$ 
goes to a pure 
$(3\,,\,3^*)\oplus (3^*\,,\,3)$, now degenerate with
the scalar meson $S$ in the same representation, 
but not with $\rho$ in 
$(8\,,\,1)\oplus (1\,,\,8)$:
\begin{eqnarray}
\mbox{(VM)}
\qquad
\left\{
\begin{array}{rcl}
\vert \pi\rangle\,, \vert \rho (\lambda=0) \rangle
 &\rightarrow& 
\vert(8\,,\,1) \oplus (1\,,\,8)\rangle\ ,
\\
\vert A_1(\lambda=0)\rangle\,, \vert s\rangle  &\rightarrow&
\vert  (3\,,\,3^\ast)\oplus(3^\ast\,,\,3)\rangle\ .
\end{array}\right.
\end{eqnarray}
Namely, the
degenerate massless $\pi$ and (longitudinal) $\rho$ at the 
phase transition point are
the chiral partners in the
representation of $(8\,,\,1)\oplus (1\,,\,8)$.

Next, we consider the helicity $\lambda=\pm1$. 
Note that
the transverse $\rho$
can belong to the representation different from the one
for the longitudinal $\rho$ ($\lambda=0$) and thus can have the
different chiral partners.
According to the analysis in Ref.~\cite{Gilman-Harari-Weinberg},
the transverse components of $\rho$ ($\lambda=\pm1$)
in the broken phase
belong to almost pure
$(3^*\,,\,3)$ ($\lambda=+1$) and $(3\,,\,3^*)$ ($\lambda=-1$)
with tiny mixing with
$(8\,,\,1)\oplus(1\,,\,8)$.
Then, it is natural to consider in VM that
they become pure $(3\,,\,3^\ast)$ and 
$(3^\ast\,,\,3)$
in the limit approaching the chiral restoration point~\cite{HY:PRep}:
\begin{eqnarray}
\vert \rho(\lambda=+1)\rangle \rightarrow 
  \vert (3^*,3)\rangle\ ,\quad
\vert \rho(\lambda=-1)\rangle \rightarrow 
  \vert (3,3^*)\rangle \ .
\end{eqnarray}
As a result,
the chiral partners of the transverse components of $\rho$ 
in the VM will be  themselves.

The formulation of the VM was first done in the large flavor 
QCD~\cite{HY:VM}, and then in the hot and dense 
QCD~\cite{HS:VM,HKR:VM}.
The formulation was done 
within the framework of the hidden local symmetry
(HLS)~\cite{BKUYY,BKY:PRep}, in which
it is possible to perform a
systematic derivative expansion 
(see Ref.~\cite{HY:PRep} for a review).

At the leading order of the chiral perturbation with HLS
the Lagrangian includes three parameters:
the pion decay constant $F_\pi$; the HLS gauge coupling $g$;
and a parameter $a$.
Using these three parameters, the $\rho$ meson mass $m_\rho$,
the $\rho$-$\gamma$ mixing strength $g_\rho$,
the $\rho$-$\pi$-$\pi$ coupling strength $g_{\rho\pi\pi}$
and
the direct $\gamma$-$\pi$-$\pi$ coupling strength $g_{\gamma\pi\pi}$
are expressed as
\begin{equation}
m_\rho^2 = g^2 a F_\pi^2 \ , \quad
g_\rho = g a F_\pi^2 \ , \quad
g_{\rho\pi\pi} = \frac{a}{2} \, g \ , \quad
g_{\gamma\pi\pi} = 1 - \frac{a}{2}
\ .
\end{equation}
{}From these expression, one can easily see that the vector dominance
(VD) of the electromagnetic form factor of the pion,
i.e. $g_{\gamma\pi\pi}=0$, is satisfied for $a = 2$.
We would like to stress that
the VD at zero temperature and density
is accidentaly satisfied:
The parameter $a$ is $4/3$ at the bare level and it becomes
$2$ in the low-energy region 
by including the quantum correction~\cite{HY:VD}.
This can be rephrased in the following way:
the parameter $a$ at the large $N_c$ limit is
$4/3$ and it becomes $2$ when the $1/N_c$ corrections are
included~\cite{HMY:AdS}.

The most important ingredient to formulate the VM in hot matter
is the intrinsic temperature dependence of the parameters
of the HLS Lagrangian~\cite{HS:VM,Sasaki:D} introduced through 
the Wilsonian matching between the HLS and QCD:
The Wilsonian matching near the critical temperature $T_c$
provides the following behavior for the bare parameters $a$ and $g$:
\begin{equation}
g(\Lambda;T) \sim \langle \bar{q}q \rangle \rightarrow 0 \ ,
\quad
a(\Lambda;T) - 1 \sim \langle \bar{q}q \rangle^2 \rightarrow 0 \ ,
\quad
\mbox{for} \ T \rightarrow T_c \ .
\label{bare g a}
\end{equation}
It was shown~\cite{HY:PRep,HS:VM,HKR:VM}
that these conditions are protected by the fixed point of 
the RGEs and never receives 
quantum corrections at the critical point.
Thus the parametric vector meson mass determined at 
the on-shell of the vector meson also vanishes since it is
proportional to the vanishing gauge coupling constant.
The vector meson mass $m_\rho$ defined as a pole position of 
the full vector meson propagator has the hadronic corrections 
through thermal loops, which are proportional to
the gauge coupling constant~\cite{HS:VM,HKR:VM,HS:VD}.
Consequently the vector meson pole mass also goes to zero
for $T \to T_c$:
\begin{equation}
m_\rho(T) \sim \langle \bar{q}q \rangle \to 0\,.
\end{equation}
We would like to stress that the VD
is strongly violated near
the critical point associated with the dropping $\rho$
in the VM in hot matter~\cite{HS:VD}:
\begin{equation}
a(T) \rightarrow 1 \ , 
\quad
\mbox{for} \ T \rightarrow T_c \ .
\end{equation}

\setcounter{equation}{0}
\section{Thermal Dilepton Spectra in the VM}
\label{sec:DS}

We should note that
the conditions in Eq.~(\ref{bare g a}) hold
{\it only in the vicinity of $T_c$}:
They are not valid any more far away from
$T_c$ where ordinary hadronic temperature
corrections are dominant.
For expressing a temperature above which the intrinsic
effect becomes important,
we introduce a temperature $T_f$, 
so-called flash temperature~\cite{BLR:flash,BLR}.
The VM and therefore the dropping $\rho$ mass become 
transparent for $T>T_f$.
On the other hand, we expect that
the intrinsic effects are negligible in the low-temperature
region below $T_f$:
Only hadronic temperature corrections are considered for $T < T_f$.
Based on the above consideration, we adopt the following
ansatz of the temperature dependences of the 
bare $g$ and $a$:~\footnote{
 As was stressed in Refs.~\cite{HY:PRep,Sasaki:D}, the VM should be
 considered only as the limit.  So we include the temperature
 dependences of the parameters only for $T_f < T < T_c - \epsilon$.
}
\begin{equation}
\left\{\begin{array}{l}
  g(\Lambda;T) = \mbox{(constant)} 
 \\
  a(\Lambda;T) - 1 = \mbox{(constant)} 
\end{array} \right\}
\ \mbox{for}\ T < T_f \ ,
\quad
\left\{\begin{array}{l}
  g(\Lambda;T) \propto \langle \bar{q}q \rangle_{T} 
 \\
  a(\Lambda;T) - 1 \propto \langle \bar{q}q \rangle_{T}^2
\end{array} \right\}
\ \mbox{for}\ T > T_f \ .
\label{bare t dep}
\end{equation}
Here we
would like to remark that the Brown-Rho scaling deals with the
quantity directly 
locked to the quark condensate and hence
{\it the scaling masses are
achieved exclusively 
by the intrinsic effect} in the present framework.

\begin{figure}
\begin{center}
\includegraphics[width = 4.8cm]{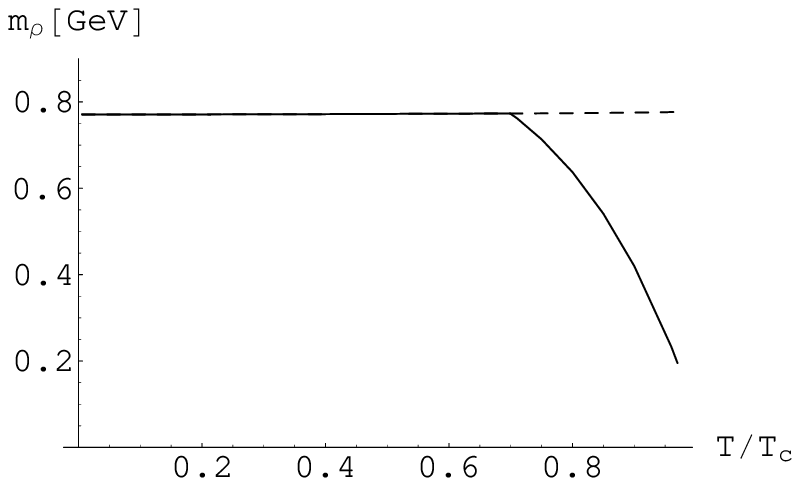}
\hspace{0.3cm}
\includegraphics[width = 4.8cm]{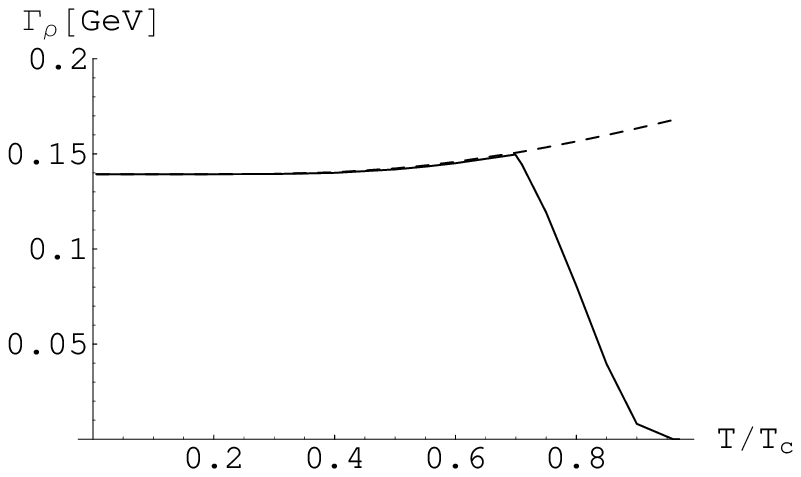}
\hspace{0.3cm}
\includegraphics[width = 4.8cm]{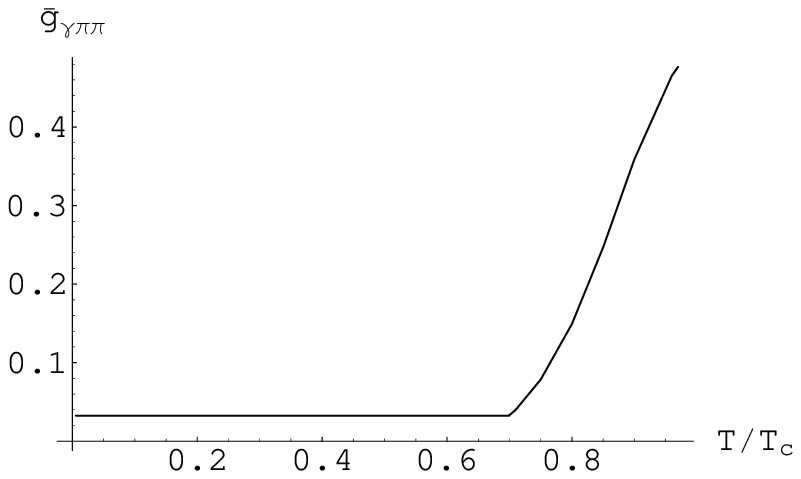}
\\
{\small
(a)
\hspace{4.5cm}
(b)
\hspace{4.5cm}
(c)
}
\end{center}
\caption[]{
Temperature dependences of (a)~the $\rho$ meson mass $m_\rho$,
(b)~the decay width $\Gamma_\rho$ and
(c)~the direct-$\gamma\pi\pi$ coupling
$\bar{g}_{\gamma\pi\pi}$.
The solid curves denote the full (both intrinsic
and hadronic) temperature dependences.
The curves with the dashed lines include only the hadronic
temperature effects. 
Note that $\bar{g}_{\gamma\pi\pi}$ includes
only the intrinsic effects.
}
\label{fig:mass width gpp}
\end{figure}

In Ref.~\cite{HS:HLSDL}, we calcuated the mass and the 
width of $\rho$ meson as well as the direct-$\gamma\pi\pi$ coupling
strength in hot matter using the in-medium parameters
determined in the above way.
We show the resultant temperature dependences
in Fig.~\ref{fig:mass width gpp}.
Figures~\ref{fig:mass width gpp}(a) and (b) show that,
below the flash temperature $T_f$, both the mass and width
slighly increase with temperature caused by the hadronic 
temperature effects.
For $T > T_f$, on the other hand, 
the intrinsic effects become dominant and both the mass and
width decrease rapidly toward zero.
Figure~\ref{fig:mass width gpp}(c)
shows that, 
in the temperature region below $T_f$,
$\bar{g}_{\gamma\pi\pi}$ is almost zero realizing the vector 
dominance (VD).
Above $T_f$ the parameter $a$ starts
to decrease from 2 to 1 due to the intrinsic effect.
This causes an increase of $\bar{g}_{\gamma\pi\pi}$ toward $1/2$,
which implies the strong violation of the VD.

\begin{figure}[htbp]
\begin{center}
\includegraphics[width = 7cm]{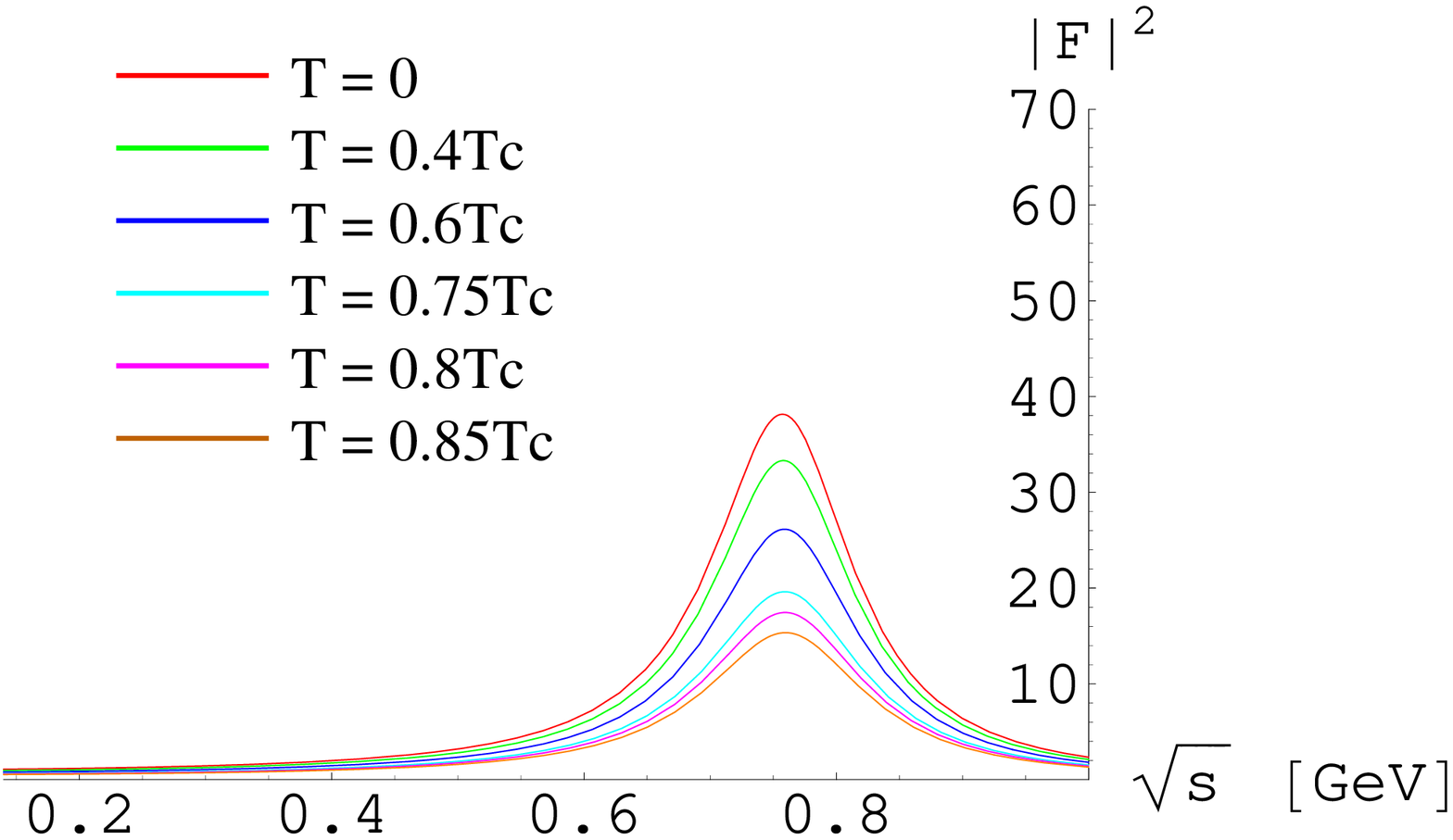}
\hspace{0.5cm}
\includegraphics[width = 7cm]{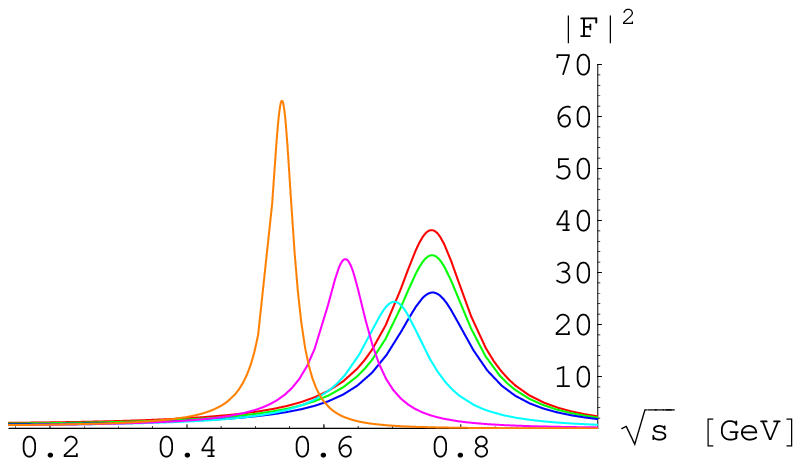}
\\
{\small (a) \hspace{7cm} (b)}
\end{center}
\caption{
Electromagnetic form factor of the pion as a function of
the invariant mass $\sqrt{s}$ for several temperatures.
The curves in the left panel (a) include only the hadronic
temperature effects and those in the right panel (b) include
both intrinsic and hadronic
temperature effects.
}
\label{fig:form}
\end{figure}

Figure~\ref{fig:form} shows the pion electromagnetic 
form factor for several temperatures.
%%%%%%%%%%%%%%%%%%%%%%%%%%%%%%%%%%%%%%%%%%%%
%%%%%%%%%%%%%%%%%%%%%%%%%%%%%%%%%%%%%%%%%%%
In Fig.~\ref{fig:form} (a) 
only the hadronic temperature corrections are included in 
the form factor.
There is no remarkable shift of the $\rho$ meson mass
but the width becomes broader with increasing temperature, 
which is consistent with the previous study~\cite{SK}.
In Fig.~\ref{fig:form} (b) the intrinsic temperature effects
are also 
included into all the parameters in the form factor.
At the temperature below $T_f$, 
the hadronic effect dominates the form factor,
so that the curves for $T = 0$, $0.4T_c$ and $0.6T_c$
agree with the corresponding ones in Fig.~\ref{fig:form}(a).
At $T = T_f$ the intrinsic effect starts to contribute
and thus in the temperature region above $T_f$ 
the peak position of the form factor moves as
$m_\rho(T) \rightarrow 0$ with increasing temperature toward $T_c$.
Associated with this dropping $\rho$ mass,
the width becomes narrow, and 
the value of the form factor at the peak
grows up as~\cite{HS:VM}
\begin{eqnarray}
\left\vert \frac{g_\rho g_{\rho\pi\pi}}{m_\rho\Gamma_\rho}
\right\vert^2 
\sim 
\left( \frac{g_\rho}{g_{\rho\pi\pi}m_\rho^2} \right)^2
\sim 
\frac{1}{g^2}
\ .
\end{eqnarray}

Now, let us show the thermal dipepton production rate predicted 
in the VM.
A lepton pair is emitted from the hot matter
through a decaying virtual photon.
The differential production rate in the medium for a fixed 
temperature $T$
is expressed in terms of the imaginary part of the photon 
self-energy $\mbox{Im}\Pi$ as
\begin{equation}
\frac{dN}{d^4q}(q_0,\vec{q};T)
=\frac{\alpha^2}{\pi^3 M^2}\frac{1}{e^{q_0/T}-1}
\mbox{Im}\Pi (q_0,\vec{q};T)\,,
\label{rate}
\end{equation}
where $\alpha = e^2/4\pi$ is the electromagnetic coupling constant,
$M$ is the invariant mass of the produced dilepton and 
$q_\mu=(q_0,\vec{q})$ denotes the momentum of the virtual photon.
We will focus on an energy region around the $\rho$ meson mass
scale in this analysis.
In this energy region it is natural to expect that
the photon self-energy is dominated by the two-pion process
and its imaginary part is related to the pion electromagnetic 
form factor ${\cal F}(s;T)$ through 
\begin{equation}
\mbox{Im}\Pi(s;T)
= \frac{1}{6\pi\sqrt{s}}
\left( \frac{s - 4m_\pi^2}{4} \right)^{3/2}
\left| {\cal F}(s;T) \right|^2\,,
\label{Im Pi}
\end{equation}
with the pion mass $m_\pi$.

\begin{figure*}
\begin{center}
\includegraphics[width = 6.5cm]{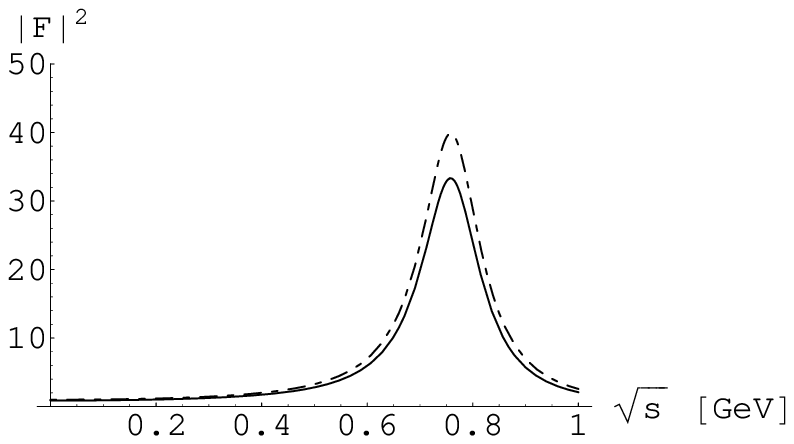}
\hspace*{0.5cm}
\includegraphics[width = 6.5cm]{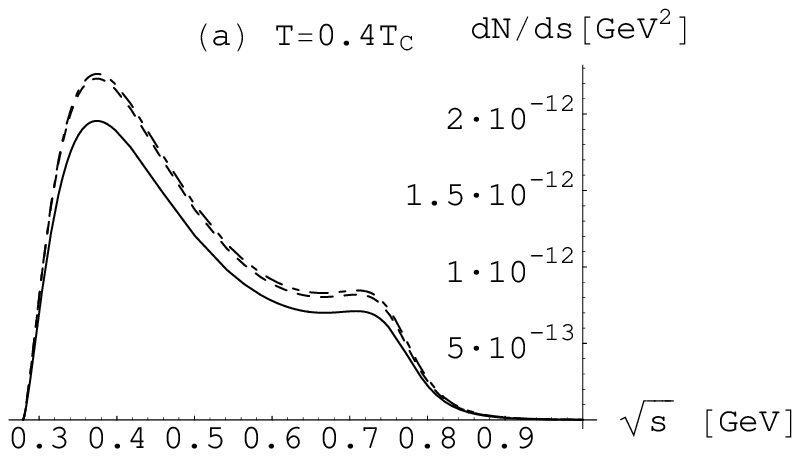}
\\
\includegraphics[width = 6.5cm]{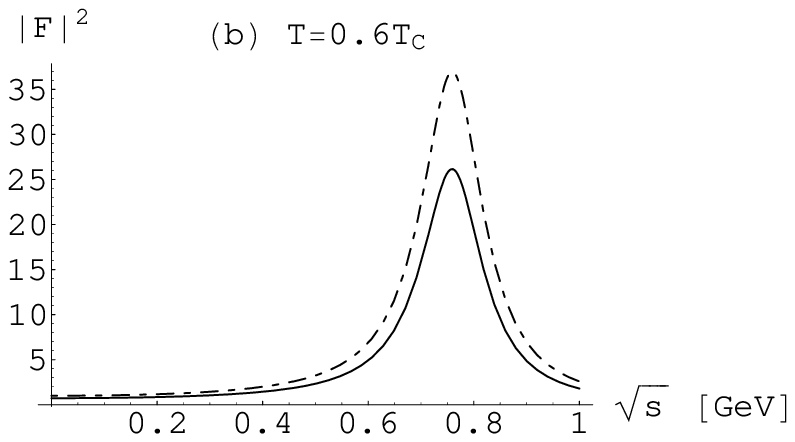}
\hspace*{0.5cm}
\includegraphics[width = 6.5cm]{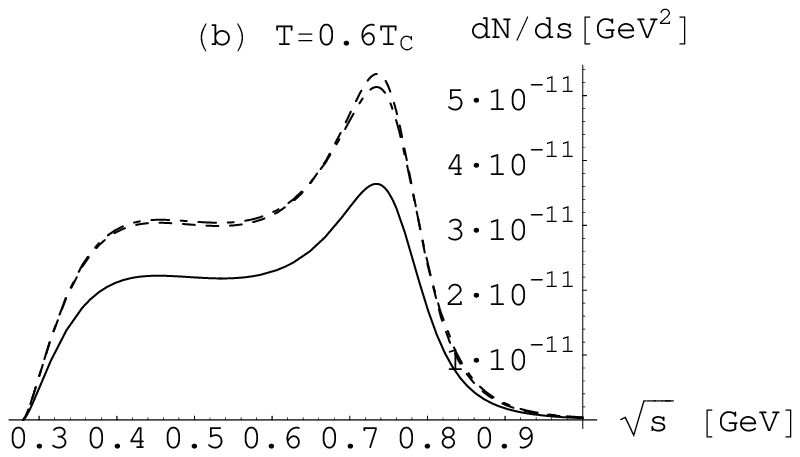}
\\
\includegraphics[width = 6.5cm]{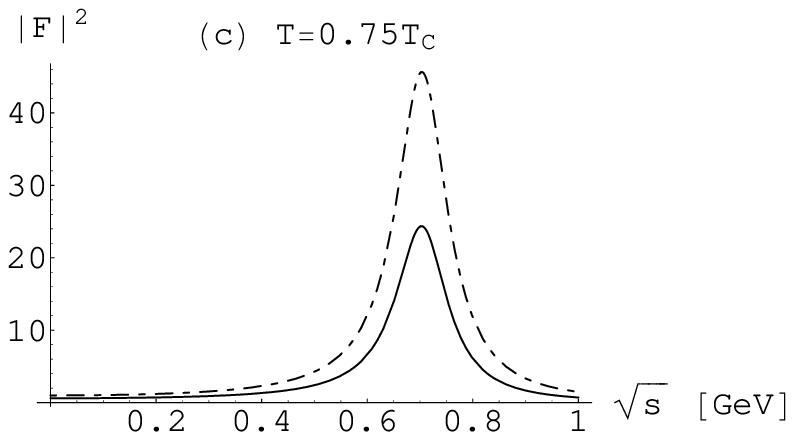}
\hspace*{0.5cm}
\includegraphics[width = 6.5cm]{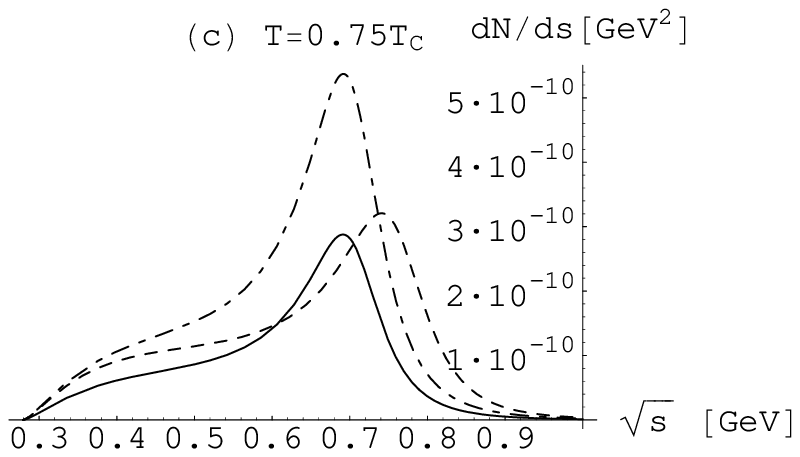}
\\
\includegraphics[width = 6.5cm]{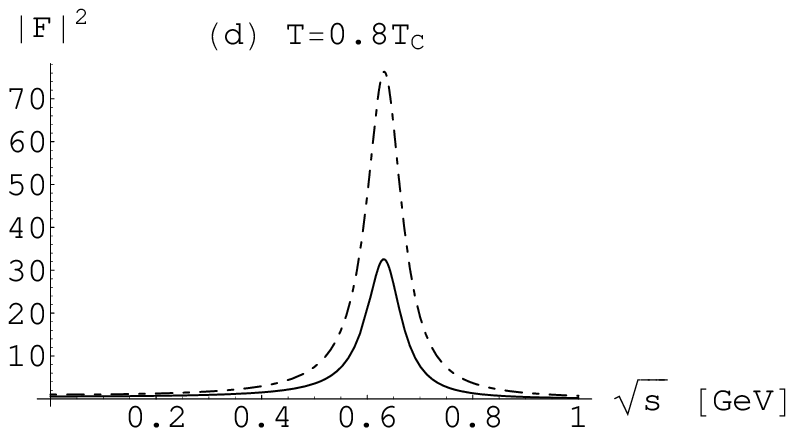}
\hspace*{0.5cm}
\includegraphics[width = 6.5cm]{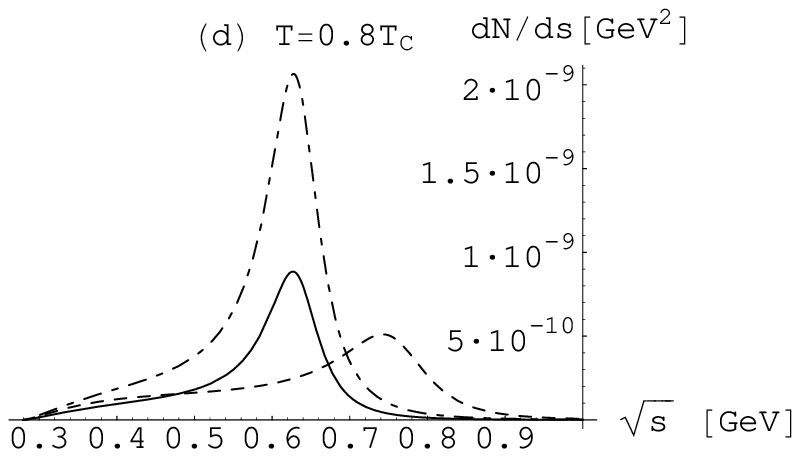}
\\
\includegraphics[width = 6.5cm]{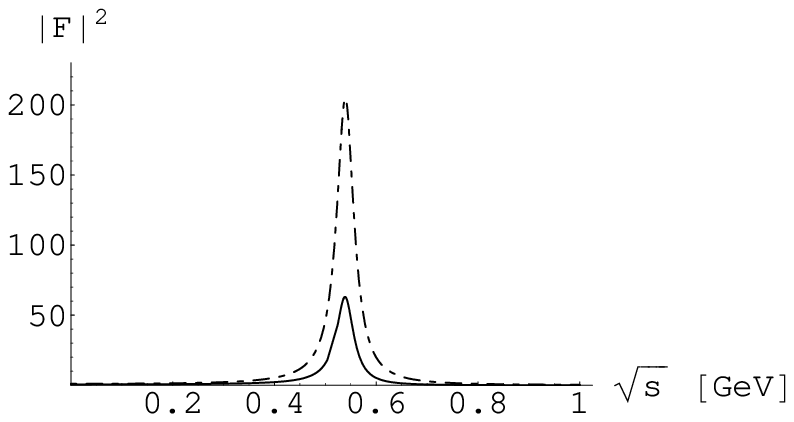}
\hspace*{0.5cm}
\includegraphics[width = 6.5cm]{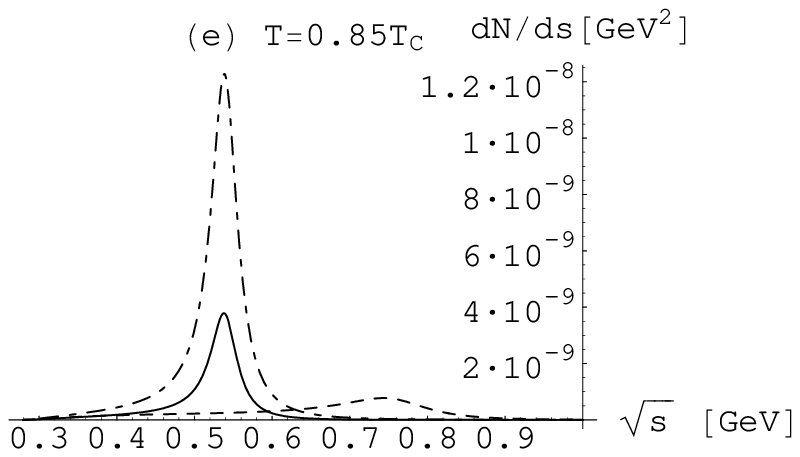}
\end{center}
\caption[]{
Electromagnetic form factor of the pion (left) and
dilepton production rate (right) as a function of the invariant 
mass $\sqrt{s}$ for various temperatures.
The solid lines include the effects of the violation of the VD.
The dashed-dotted lines correspond to the analysis assuming
the VD. In the dashed curves in the right-hand figures, 
the parameters at zero temperature were used.
}
\label{fig:dl}
\end{figure*}

As noted,
the vector dominance (VD)
is controlled by the parameter $a$ in the HLS theory.
The VM leads to the strong violation of the VD 
(indicated by ``$\Slash{\rm VD}$'') 
near the chiral symmetry restoration point, which can be traced 
through the Wilsonian matching and the RG evolutions.
Thus the direct photon-$\pi$-$\pi$ $g_{\gamma\pi\pi}$ coupling
yields 
non-vanishing contribution to the form factor together with the 
$\rho$-meson exchange.
In Ref.~\cite{HS:HLSDL},
we compared the dilepton spectra predicted in the VM
(including the effect of $\Slash{\rm VD}$) with those
obtained by assuming the VD, i.e. taking 
$g_{\gamma\pi\pi}=0$.
Figure~\ref{fig:dl} shows the form factor and the dilepton 
production rate integrated over three-momentum,
in which the results with VD and $\Slash{\rm VD}$
were compared.
%%%%%%%%%%%%%%%%%%%%%%%%%%%%%%%%%%%%%%%%%%%
%%%%%%%%%%%%%%%%%%%%%%%%%%%%%%%%%%%%%%%%%%%%%%%
The figure shows a clear difference between the curves
with VD and $\Slash{\rm VD}$.
In the low-temperature region
$T \ll T_f$,
the hadronic effects are dominant compared with the intrinsic ones, 
so both curves almost coincide.
A difference between them starts to appear around $T=T_f$
and increases with 
temperature.
It can be easily seen that the $\Slash{\rm VD}$ gives a reduction
compared to the case with keeping the VD.
The features of the form factor as well as the dilepton production 
rate coming from two-pion annihilation shown in 
Fig.~\ref{fig:dl}(a)-(e) are summarized below for each temperature:
%%%%%%%%%%%%%%%%%%%%%%%%%%%%%%%%%%%
\begin{description}
\item[(a) and (b) (below $T_f$) :]
The form factor, which has a peak at the $\rho$ meson mass
$\sqrt{s} \sim 770$ MeV, is slightly suppressed with 
increasing temperature.
An extent of the suppression in case with $\Slash{\rm VD}$ is 
greater than that with VD.
This is due to decreasing of the $\rho$-$\gamma$ mixing strength
$g_\rho$ at finite temperature.
At $T < T_f$, $g_\rho$ mainly decreases by hadronic corrections.
In case with VD, however, $g_\rho$ is almost constant.
The dilepton rate (a) has two peaks, one is at the $\rho$ meson mass
and another one is lying around low-mass region.
The later peak comes from the Boltzman factor of Eq.~(\ref{rate}).
For a rather low-temperature the production rate is much enhanced
compared with the $\rho$ meson peak
since the yield in the higher-mass region is suppressed by 
the statistical factor.
With increasing temperature those peaks of the production rate (b)
are enhanced and the peak at $\sqrt{s} \sim m_\rho$ clearly appears.
In association with decreasing $g_\rho$, one sees a reduction
of the dilepton rate with $\Slash{\rm VD}$.
%%%%%%%%%%%%%%%%%%%%%%%%%%%%%%%%%%%%%%%%
\item[(c), (d) and (e) (above $T_f$) :]
Since the intrinsic temperature effects are turned on, a shift of
the $\rho$ meson mass to lower-mass region can be seen.
Furthermore, 
the form factor, which becomes narrower with increasing
temperature due to the dropping $m_\rho$, exhibits an obvious
discrepancy between the cases with VD and $\Slash{\rm VD}$.
The production rate based on the VM 
(i.e., the case with $\Slash{\rm VD}$) is suppressed compared
to that with the VD.
We observe that the suppression is more transparent 
for larger temperature:
The suppression factor is $\sim 1.8$ in (c), $\sim 2$ in (d)
and $\sim 3.3$ in (e). 

As one can see in (c), the peak value of the rate
predicted by the VM
in the temperature region slightly above the flash temperature
is even smaller than the one obtained by the vacuum parameters,
and the shapes of them are quite similar to each other.
This indicates that it might be difficult to measure the 
signal of the dropping $\rho$ experimentally, if this
temperature region is dominant in the evolution of the fireball.
In the case shown in (d), on the other hand,
the rate by the VM 
is enhanced by a factor of 
about two compared with the one by the vacuum $\rho$.
The enhancement becomes prominent near the critical temperature
as seen in (e).
These imply that we may have a chance to discriminate the
dropping $\rho$ from the vacuum $\rho$.

\end{description}

\setcounter{equation}{0}
\section{Summary and Discussions}
\label{sec:sum}

We studied the pion electromagnetic form factor and the thermal
dilepton production rate from the two-pion annihilation
within the hidden local symmetry (HLS) theory as an effective
field theory of low-energy QCD.
In the HLS theory the chiral symmetry is restored as the vector
manifestation (VM) in which the massless $\rho$ meson joins
the same chiral multiplet as pions.
In order to determine the temperature dependences of the
parameters of the HLS Lagrangian, the Wilsonian matching to
the operator product expansion at finite temperature was made
by applying the matching scheme developed 
in the vacuum~\cite{HY:WM,HY:PRep} and at the critical 
temperature~\cite{HS:VM,HKRS:SUS,HS:VD,PiV}.

In the notion of the Wilsonian matching to define a bare theory
in hot environment,
the bare parameters are dependent on temperature, which
are referred as the intrinsic temperature effects.
At low temperatures the chiral properties of in-medium hadrons
are dominated by ordinary hadronic loop corrections.
The dropping $\rho$ is realized in the HLS framework due to
the intrinsic effects and thus they play crucial roles especially
near the chiral phase transition.
In order to see an influence of the intrinsic temperature effects,
we presented the form factor including full temperature effects,
i.e., the intrinsic and hadronic effects, and compared
with that including only hadronic corrections.
The $\rho$ meson mass $m_\rho$ is almost stable against
the hadronic corrections and one does not obtain the dropping $m_\rho$.
Accordingly the peak of the form factor including only the 
hadronic effects is located at around $\sqrt{s} \sim m_\rho \sim 770$
MeV even at finite temperature.
The form factor is reduced with increasing temperature and
correspondingly becomes broader.
On the other hand, the Wilsonian matching procedure certainly 
involves the intrinsic temperature effects in the analysis
and provides the dropping $m_\rho$ as the VM.
The form factor {\it above the flash temperature $T_f$} thus 
starts to present a shift of $m_\rho$ to lower invariant mass region.
Associated with the dropping $\rho$, the form factor becomes sharp.

One of the significant predictions of the VM is a strong violation 
of the vector dominance (VD) of the pion form factor.
The VM predicts that the VD is violated near the transition 
temperature $T_c$ in which the direct photon-$\pi$-$\pi$ coupling
does contribute to the form factor in addition to the $\rho$-meson
mediation.
It crucially affects the analysis of dilepton yields.
We presented the form factor and the dilepton production rate
with and without the VD assumption together with the dropping $\rho$.
For $T \ll T_f$ the result shows only 
a small difference between those
two cases since the VD is still well satisfied in low temperatures.
A clear difference can be seen for $T > T_f$ where the intrinsic
temperature effects contribute to the physical quantities.
The form factor and consequently the dilepton production rate
with taking account of the violated VD are reduced and 
exhibit an obvious difference near $T_c$ compared to those 
with the VD.

Several comments are in order:

The HLS Lagrangian has only pions and vector mesons as physical
degrees of freedom, and a time evolution was not considered
in this work.
Thus it is not possible to make a direct comparison of our results
with experimental data.
However a {\it naive} dropping $m_\rho$ formula, i.e., $T_f = 0$, 
as well as VD in hot/dense matter are sometimes used for
theoretical implications of the data.
As we have shown in this paper, the intrinsic temperature effects
together with the violation of the VD give a clear difference
from the results without including those effects.
It may be then expected that a field theoretical analysis
of the dropping $\rho$ as presented in this work and a reliable
comparison with dilepton measurements will provide an evidence
for the in-medium hadronic properties associated with the chiral
symmetry restoration, if complicated hadronization processes do
not wash out those changes.

Recently the chiral perturbation theory with including vector
and axial-vector mesons as well as pions has been constructed%
~\cite{HS:GHLS,hidaka} based on the generalized HLS%
~\cite{BKY:NPB,BKY:PRep,BFY:GHLS}.
In this theory the dropping $\rho$ and $A_1$ meson masses
were formulated and it was shown that the dropping masses
are related to the fixed points of the RGEs which gives
a VM-type restoration and that the VD is strongly violated
also in this case.
Inclusion of the effect of $A_1$ meson as well as
the effect of collisional broadening will be done in future
work~\cite{HS:DL2}.

%%%%%%%%%%%%%%%%%%%%%%%%%%%%%%%%%%%%%%%%%%%%%%%%%%%%%
%%%%%%%%%%%%%%%%%%%%%%%%%%%%%%%%%%%%%%%%%%%%%%%%%%%%%%

\subsection*{Acknowledgments}

We are grateful to Gerry Brown, Bengt Friman and Mannque Rho 
for fruitful discussions and comments.
We also thank Jochen Wambach for stimulating discussions.
The work of C.S. was supported in part by the Virtual Institute
of the Helmholtz Association under the grant No. VH-VI-041.
The work of M.H. 
is supported in part by the Daiko Foundation \#9099 and
the 21st Century
COE Program of Nagoya University provided by Japan Society for the
Promotion of Science (15COEG01).

%%%%%%%%%%%%%%%%%%%%%%%%%%%%%%%%%%%%%%%%%%%%%%%%%
%%%%%%%%%%%%%%%%%%%%%%%%%%%%%%%%%%%%%%%%%%%%%%%%%%

%\renewcommand{\baselinestretch}{0.6}

\end{document}